\documentclass[pra,final,twocolumn,amsmath,amssymb,superscriptaddress,amsfonts]{revtex4-1}
\usepackage{graphicx,graphics}
\usepackage{color}
\usepackage{gensymb}
\usepackage{epsfig}
\usepackage{amsmath}
\usepackage{amssymb}
\usepackage{bm}

\usepackage{textcomp}

\begin{document}

\title{      Orbital Symmetry and Orbital Excitations in High-$T_c$ Superconductors}

\author{     Andrzej M. Ole\'s  }
\email{a.m.oles@fkf.mpi.de, corresponding author}
\affiliation{Max Planck Institute for Solid State Research,
             Heisenbergstrasse 1, D-70569 Stuttgart, Germany }
\affiliation{\mbox{Marian Smoluchowski Institute of Physics, Jagiellonian University,
             Prof. S. \L{}ojasiewicza 11, PL-30348 Krak\'ow, Poland}}
             
\author{     Krzysztof Wohlfeld  }
\affiliation{\mbox{Institute of Theoretical Physics, Faculty of Physics, University of Warsaw, 
             Pasteura 5, PL-02093 Warsaw, Poland}}
             
\author{     Giniyat Khaliullin }
\affiliation{Max Planck Institute for Solid State Research,
             Heisenbergstrasse 1, D-70569 Stuttgart, Germany }

\date{5 May, 2019}

\begin{abstract}
We discuss a few possibilities of high-$T_c$ superconductivity with
more than one orbital symmetry contributing to the pairing. First, we
show that the high energies of orbital excitations in various cuprates
suggest a simplified model with a single orbital of $x^2-y^2$
symmetry doped by holes. Next, several routes towards involving both $e_g$
orbital symmetries for doped holes are discussed:
(i) some give superconductivity in a CuO$_2$ monolayer on Bi2212
superconductors, Sr$_2$CuO$_{4-\delta}$, Ba$_2$CuO$_{4-\delta}$, while
(ii) others as nickelate heterostructures or Eu$_{2-x}$Sr$_x$NiO$_4$,
could in principle realize it as well.
At low electron filling of Ru ions, spin-orbital
entangled states of $t_{2g}$ symmetry contribute in Sr$_2$RuO$_4$.
Finally, electrons with both $t_{2g}$ and $e_g$ orbital symmetries
contribute to the superconducting properties and nematicity of
Fe-based superconductors, pnictides or FeSe.
Some of them provide examples of orbital-selective Cooper pairing.\\
\textit{Published in:}  
Special Issue \textit{"From Cuprates to Room Temperature Superconductors"} 
dedicated to the anniversary of Professor K. Alex M\"uller; 
Condensed Matter \textbf{4}, 46 (2019).
\end{abstract}

\maketitle

\section{Introduction: towards superconductivity with orbital degrees of freedom}
\label{sec:intro}

Significance of the discovery of high-$T_c$ superconductivity by
Bednorz and Muller \cite{Bed86} for recent progress in the quantum
many-body theory cannot be overestimated --- it triggered a huge
amount of innovative research on quantum materials and unconventional
superconductors, both in the experiment and in the theory.
In spite of reaching a qualitative understanding of the nature of the
superconducting (SC) state in different transition metal compounds,
several open problems remain. Some are related to cuprates, as the
astonishing complexity of the phase diagram and the physical origin of
the temperature $T^*$ observed well above $T_c$ itself
\cite{Lee06,Oga08,Voj09,Bia12,Kei15}; others are more general and
include questions about the origin of pairing
\cite{Pag10,Hir11,Sca12,Si16,Fer17}, optimal conditions for the onset
of the SC state \cite{Cav94,Kas98,Ima98,Mac03}, variation under
pressure \cite{Krz16}, and the actual orbital symmetry at the Fermi
surface \cite{Gra09}.

In this short review we concentrate on the last question.
After the discovery of high-$T_c$ superconductivity in cuprates it was
believed that lifting of degeneracy of $e_g$ orbitals was important to
obtain the SC state with a high value of $T_c$ \cite{Oht91,Pav01,Sak10}.
Indeed, typically in cuprates degeneracy is quite lifted and relevance
of Jahn-Teller effect is controversial. It is under discussion whether
the effective model for cuprates has nondegenerate orbitals and may be
represented by extended Hubbard model with on-site Coulomb and further
neighbor hopping \cite{Fei96}. Yet, it is derived for the case of large
splitting of $e_g$ orbitals, while orbital degrees of freedom might
play an important role in the SC instability and the
multiorbital Hubbard model is a standard model for all high-$T_c$
superconductors in general, where the dome of $T_c$ occurs by driving
the chemical potential in the proximity of a Lifshitz transition
\cite{Bia18}.

We shall not address here the role played by electron-phonon coupling
which is expected to contribute in cuprates and is a driving force of
recently discovered superconductivity in H$_3$S \cite{Dro15,Cap17}. In
particular, since the degenerate orbital degrees of freedom necessarily
come along with strong Jahn-Teller coupling \cite{Mul99,Kel08},
the electron-phonon coupling seems to be essential for SC instabilities
in all systems with orbital degeneracy. Nevertheless, to make this
review more focused, we shall limit ourselves to the consequences of
the orbital degeneracy for the models containing solely electronic
degrees of freedom --- thus leaving the interplay of the
electron-phonon coupling and the orbital degeneracy in the high-$T_c$
superconductors for another work.

The outline of this paper is as follows. One early idea in the theory
of cuprate superconductivity was that orbital excitations could
contribute to the pairing mechanism, as discussed also in section 2.1,
but this was not supported by more recent developments. A usual
situation is that the pairing occurs for holes in a single molecular
orbital of $x^2-y^2$ symmetry, see section 2.2.
But certainly an interesting question is whether allowing for the
presence of holes in both $e_g$ orbitals would not lead to enhanced SC
instabilities. This idea has its roots in the Jahn-Teller physics in
cuprates \cite{Mul99,Kel08}, as well as in the observation that the
propagation of a hole in a Mott (or charge-transfer) insulator is much
richer when both $e_g$ orbitals can participate \cite{Zaa93}.
Partial filling of both $e_g$ orbitals could be realized in the highly
overdoped CuO$_2$ monolayer grown on Bi$_2$Sr$_2$CaCu$_2$O$_{8+\delta}$
(Bi2212) \cite{Zho16,Jia18}. We remark that the symmetry of the SC
phase in Bi2212 has been extensively discussed in the literature
\mbox{(see, e.g. \cite{Li99,Mis02,Hoo03,Lat04,Klemm,Zhu19})}.

Furthermore, we emphasize that two-dimensional (2D) systems are special,
and possible SC instability was predicted for a layered geometry
of NiO$_2$ planes in LaNiO$_3$/LaMO$_3$ superlattices \cite{Cha08}.
We follow this idea and discuss briefly remarkable similarity between
overdoped cuprates and nickelates in sections 3.1 and 3.2.
Furthermore, superconductivity occurs also in metallic systems with
$t_{2g}$ degrees of freedom, i.e., in planar ruthenate Sr$_2$RuO$_4$
(section 3.3) and in Fe-based superconductors (section 3.4).
In the latter systems orbital fluctuations are expected to contribute
\cite{Hir11,Kno11}. The former systems are of particular interest as
there spin-orbit interaction entangles spin-orbital degrees of freedom
and the orbital states become mixed \cite{Vee14}. A planar iridate
Sr$_2$IrO$_4$ with an even stronger spin-orbit coupling shows much of
the cuprate phenomenology \cite{Ber19}, but no superconductivity was
reported so far. This review is summarized in section 4.

\section{The role of orbitals in superconducting cuprates}
\label{sec:exci}

\subsection{Earlier theoretical proposals}
\label{sec:specu}

Following the idea of Jahn-Teller physics in cuprates
\cite{Mul99,Kel08}, a question arises how many orbital symmetries
should be included in a minimal realistic model for cuprates.
Already in the early years of high-$T_c$, the idea of going beyond the
single band picture by including O($2p$) orbitals in the three-band
model has emerged \cite{Var87,Eme87,Ole87}.
Whether or not the oxygen orbitals could be fully integrated out is
still not fully resolved \cite{Arr09,Han10,Ebr14}.
Leaving this issue open, we shall present here an overview of the
role played by the copper orbital degrees of freedom.

It has been discussed early on that multiple orbitals contribute to
the physical properties of YBa$_2$Cu$_3$O$_{7-x}$
\cite{Biapc,Bia88,Nuc95} and La$_{2-x}$Sr$_x$CuO$_4$
\cite{Pompa,Che92}. The coupling to the lattice was employed as
sensitive to the orbital content of wave-functions ---
multiorbital components were deduced from uniaxial and hydrostatic
pressure effects on the value of $T_c$ \cite{Sak12}, and from the
effect of rhombic distortion on the polarized x-ray absorption spectra
in high-$T_c$ superconductors \cite{Sei90}.

Further research on stripes and electron-lattice interactions
suggested the presence of pseudo-Jahn-Teller effect in cuprates
\cite{Bia98,Bia00,Bia00jp,Bia01}. These phenomena follow in a natural
way from orbital pseudo-degeneracy \cite{Ber13}, and these ideas were
further developed in \cite{Mul00,BuHol,Mul14,Zho95,Ber97}. Support for
the relevance of electron-phonon interaction comes from the observation
of the isotope effect on the pseudogap temperature $T^*$ \cite{Lan99},
from strong renormalization of certain phonons by doped holes
\cite{McQ99}, and from recently observed phonon
anomalies in charge density wave states in
\mbox{cuprates \cite{Tacon,Mia18}}.

Already shortly after the discovery of high-$T_c$ cuprates,
it was suggested by Weber \cite{Web88} that
an orbital excitation could be responsible for the pairing.
In a typical copper oxide the
nearest neighbor Coulomb interaction between holes in the oxygen $p$
orbitals and the copper $3z^2-r^2$ orbital is substantially smaller
than between holes in the oxygen $p$ orbitals and the Cu $x^2-y^2$
orbital --- such an aspherical Coulomb interaction is estimated to be
of the order of 0.3-0.5 eV in the cuprates \cite{Web88}. Consequently,
an excitonic pairing mechanism was proposed: two oxygen holes can gain
energy provided that the first one excites the Cu $d$ hole from the
$x^2-y^2$ to the $3z^2-r^2$ orbital and the second one follows,
forming a pair. This mechanism was later further improved by Jarrell,
Cox, and others \cite{Jar88,Cox89} by including the superexchange
processes between the nearest neighbor oxygen $p$ orbitals and both Cu
$e_g$ orbitals. This latter mechanism was estimated to roughly triple
the strength of the coupling between the orbital exciton and the holes
on oxygen.

While the above proposal is appealing, unfortunately it is not very
realistic: the crucial role played by the copper spins is completely
neglected and it is implicitly assumed that the doped holes go to the
$\pi$-bonding oxygen orbitals. A more realistic two-orbital $e_g$
model was proposed in \cite{Zaa93}, and the onset of superconductivity
in the various versions of the two-orbital Hubbard models was studied
in more detail, e.g. in \cite{Fei92,Bud94,Buc95,Sak14}. Nevertheless,
in order to verify whether these models could be relevant to the
cuprate superconductivity, it is crucial to include the size of the
crystal-field splitting between the Cu $3z^2-r^2$ and $x^2-y^2$
orbitals --- as discussed in the next subsection.

\subsection{Orbital excitations in cuprates}
\label{sec:ctm}

In a `typical' high-$T_c$ cuprate, the CuO$_6$ octahedra are elongated
along the $c$ axis due to apical oxygen displacements,
and the degeneracy of $e_g$ orbitals is lifted.
It has been established that both electron-doped and hole-doped copper
oxides are strongly correlated electron systems in the vicinity of the
metal to charge-transfer insulator transition \cite{Web10}.
One also finds large splitting of the copper $3d$ states and only
a single Cu($3d$)--O($2p$) hybridized band crosses the Fermi surface in
doped systems. This band has the $x^2-y^2$ symmetry and becomes
half-filled in the undoped charge-transfer insulator La$_2$CuO$_4$. The
superexchange stabilizes then antiferromagnetic (AF) order \cite{Zaa88}
in this and other compounds of the cuprate family \cite{Lee06,Oga08}.
Doping generates a Fermi surface originating from a single band made of
oxygen $2p$ and copper $3d$ states, for small hole doping $x<0.3$,
as discussed by Zhang and Rice \cite{Zha88}, see Fig. 1.
In this regime $d$-wave SC phase is found, and a high value of $T_c$ is
obtained when the orbital splitting is large \cite{Oht91,Pav01,Sak10}.

\begin{figure*}[t!]
\centering
\includegraphics[width=14.5cm]{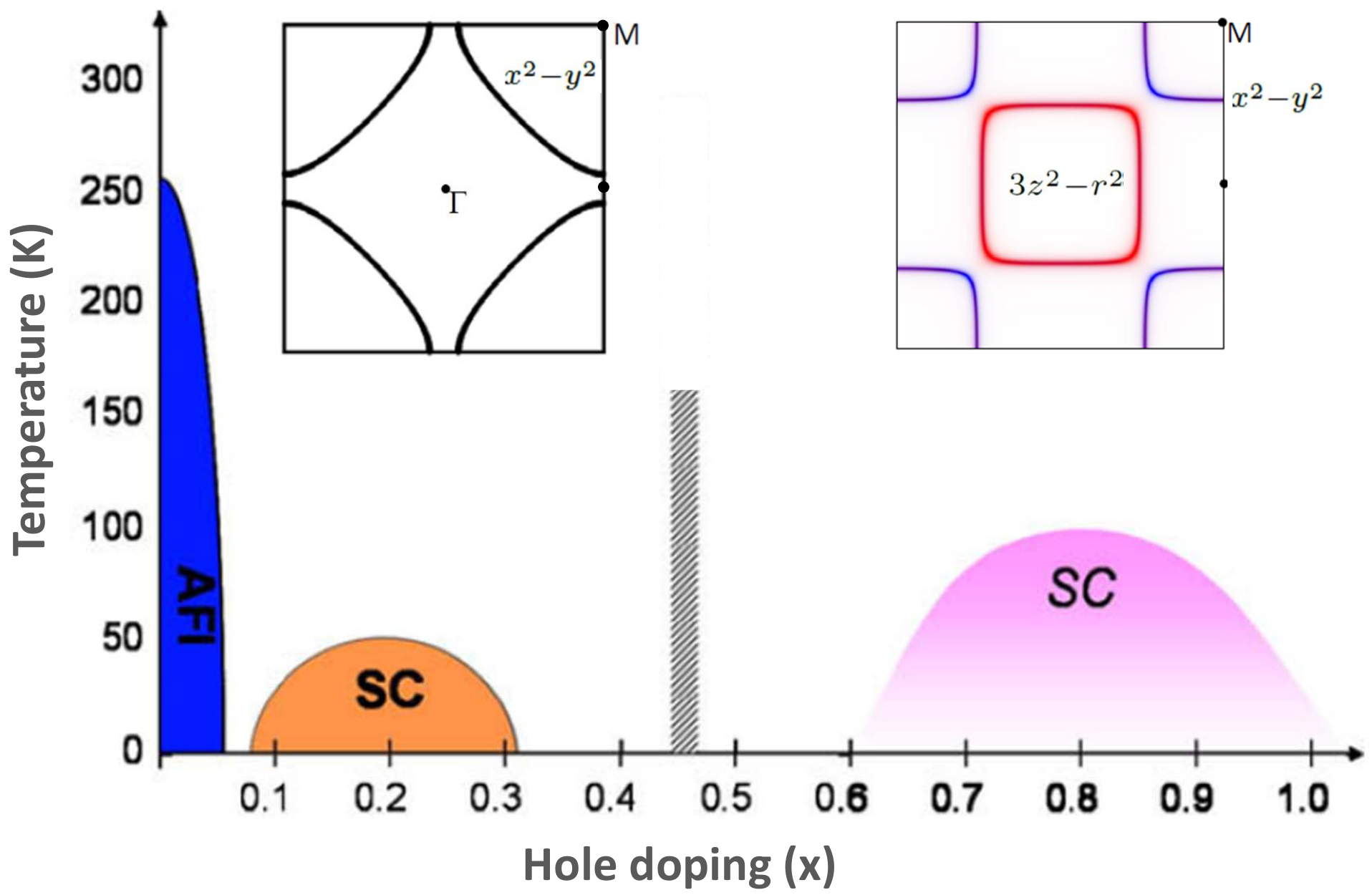}
\caption{Schematic phase diagram for cuprates as a function of hole
doping $x$, showing the route from the single-band SC phase with holes
in $x^2-y^2$-type molecular orbitals, realized in bulk cuprates (left),
to the two-orbital nodeless SC phase in the hole-rich CuO$_2$/Bi2212
monolayer (right). The insets show the corresponding Fermi surface for
one-orbital (left) and two-orbital $\{x^2-y^2,3z^2-r^2\}$ (right) SC
phase. Image is reproduced from \cite{Geb09}; right inset
for the hole-rich CuO$_2$ monolayer is reproduced from \cite{Mai18}.}
\label{fig:Cu}
\end{figure*}

Interestingly, it has taken almost 20 years both for the experiment
and the theory to unequivocally agree on the energies of the (local)
orbital excitations on the copper ion of the undoped cuprates
("$d-d$ excitations" \cite{Zaa93}). It has long been believed that,
while the ground state orbital is of ${x^2-y^2}$ character, the energy
of the ${3z^2-r^2}$ orbital excitation is rather low, for instance of
the order of $\sim 0.4$ eV --- as suggested by the optical and resonant
inelastic x-ray scattering (RIXS) measurements of La$_2$CuO$_4$
\cite{Per93,Ghi04} as well as by optical absorption for
Sr$_2$CuO$_2$Cl$_2$ \cite{Per93}. The latter result is particularly
striking, for the most recent understanding (see Table I and discussion
below) is that the ${3z^2-r^2}$ orbital excitation is not even the
lowest lying orbital excitation in Sr$_2$CuO$_2$Cl$_2$. Though, even
earlier the above result was challenged by Lorenzana and Sawatzky
\cite{Lor95}, suggesting a different interpretation of the optical
absorption spectra, in which the 0.4 eV feature was assigned to the
magnetic excitations. Finally, these results did not agree with the
theoretical estimates giving ca. 0.9 eV energy for such an orbital
excitation, according to the density-functional theory \cite{Sak10},
or with the x-ray absorption spectroscopy results showing a very weak
$3z^2-r^2$ character in the ground state of the hole-doped cuprates
\cite{Che92}.

\begin{table}[b!]
\caption{The energies of orbital excitations in various undoped
cuprates, as obtained form the RIXS experiment \cite{Mor11}
(quantum-chemistry calculations \cite{Hoz11}). Note that a classical
magnetic exchange energy $2J\simeq 0.26$ eV is subtracted from all
given energy values. Table adopted from \cite{Hoz11}.}
 \centering
\begin{ruledtabular}
 \begin{tabular}{cccc}
  \textbf{Cu($3d$) orbital}    & \textbf{La$_2$CuO$_4$} &
 \textbf{Sr$_2$CuO$_2$Cl$_2$}  & \textbf{CaCuO$_2$}     \\
 \hline
 ${3z^2-r^2}$ & 1.44 (1.37) & 1.71 (1.75) & 2.39 (2.38) \\
 ${xy}$       & 1.54 (1.43) & 1.24 (1.16) & 1.38 (1.36) \\
 ${xz/yz}$    & 1.86 (1.78) & 1.58 (1.69) & 1.69 (2.02) \\
\end{tabular}
\end{ruledtabular}
\label{tab}
\end{table}

The recent years have, however, lead to substantial advancements both
in the resolution of the RIXS experiments of the cuprates \cite{Ame11}
as well as \textit{ab-initio} quantum chemistry calculations
\cite{Hoz11}, and allowed for obtaining the results for cuprates
\cite{Hoz11,Mor11} in remarkable agreement between the theory and
experiment, as presented in Table I. We emphasize that nowhere one can
find the above-mentioned low values of the orbital excitation energies,
for all excitations have their energies substantially above 1 eV.
Finally, it is only in La$_2$CuO$_4$ that the lowest energy orbital
excitation has a $3z^2-r^2$ character; otherwise this excitation has
the {\it highest} energy. In general, a more detailed study of various
other copper oxides, \mbox{with / without} apical ligands, suggests
that the energy of the $3z^2-r^2$ orbital correlates with the
out-of-plane Cu-ligand distance $h$, although this relation is rather
complex (cf. Fig.~1 of \cite{Hoz11}).

\begin{figure*}[t!]
\centering
\includegraphics[width=14.5cm]{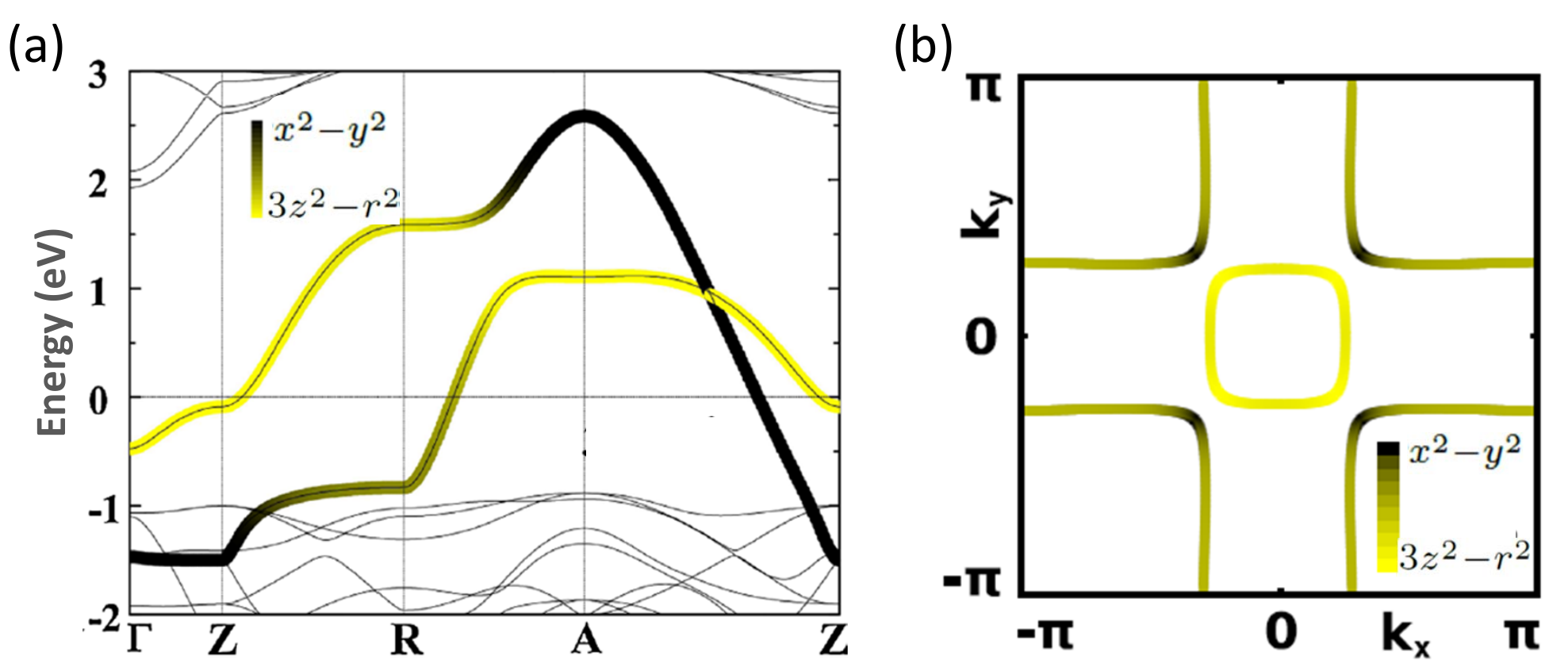}
\caption{Electronic structure for the LaNiO$_3$/LaAlO$_3$
heterostructure without strain obtained in the LDA:
(\textbf{a}) bands of $e_g$ symmetry (black to yellow shading for
orbital character), and
(\textbf{b}) corresponding cross section of the Fermi surface with the
$k_z=0$ plane. Images are reproduced from \cite{Han09}.}
\label{fig:Ni}
\end{figure*}

This situation is somewhat more subtle in doped cuprates. Again, in the
early years of high-$T_c$ research, it was expected that the occupancy
of the low-lying orbital $3z^2-r^2$ would increase upon doping
\cite{Bia88,Bia89,Nuc89,Kho91}.
However, as just discussed, in `most' of the cuprates this orbital turns
out to be the highest lying one in the undoped crystals, such a scenario
now seems to be no longer relevant. Instead, (typically) the lowest
lying $xy$ orbital either hardens by ca. 50 meV~\cite{Ell15}, or softens
by 150 meV~\cite{Fum19} with doping, depending on the compound.
Moreover, in both cases the lowest lying $d-d$ excitation has an
energy by ca. 1.5 eV higher than the ground state orbital even in
SC samples \cite{Kan19}, in agreement with Table~I.

The above discussion shows that the orbital excitations in the cuprates
have relatively high energies, suggesting that the electrons close to
the Fermi surface are clearly of a single $3d$-band character.
Surprisingly, however, some important signatures of the so-called
\textit{orbital physics}
\cite{Kug82,Fei97,Tok00,Ole05,Kha05,Nor08,Karlo,Sch12,Bis15,AMO12,Woh11,Woh13,Brz15,Brz19}
are visible there: clear experimental signatures of the collective
orbital excitation, the orbiton, have been observed in the
quasi-one-dimensional (quasi-1D) copper oxides \cite{Sch12,Bis15}.
To a large extent this turned out to be possible due to the strong
crystal-field splitting of the orbital excitations \cite{Woh11,Woh13}.
The search for similar phenomena in the quasi-2D (i.e., high-$T_c$)
cuprates is ongoing~\cite{Comin}.

\section{Superconductivity with orbital degrees of freedom}
\label{sec:orbi}

\subsection{Beyond the cuprates with the pairing in a single $x^2-y^2$ orbital}
\label{sec:Cu}

A way to enhance the value of $T_c$ in compounds isostructural with
La$_2$CuO$_4$ was found by Uchida {\it et al.} \cite{Uch06}: higher
hole doping in CuO$_2$ planes is realized by replacing La ions by Sr
ions in Sr$_2$CuO$_{4-\delta}$. The SC transition temperature $T_c=95$
K is then almost doubled for the hole doping $x\simeq 0.8$, see Fig.~1.
In this regime another nodeless SC phase arises, and the symmetry
in the orbital space is partly restored. In fact, also the band of
$3z^2-r^2$ symmetry is partly filled, and the pairing occurs jointly
for the B$_{1g}$ and A$_{1g}$ channels in Sr$_2$CuO$_{4-\delta}$ and
also in Ba$_2$CuO$_{4-\delta}$ \cite{Mai18}. This possibility was
discussed a decade ago \cite{Geb09}, and was realized,
\textit{inter alia}, in a recently discovered
superconductor with $T_c>70$~K \cite{Uch18}. The CuO$_6$ octahedra are
compressed here and $3z^2-r^2$ orbitals contribute at the Fermi surface.
The pairing strength is large relative to a similar calculation for an
optimally doped Hubbard model. The theory predicts that the pairing
strength depends on the shape of the Fermi surface \cite{Mai18},
see the right inset of Fig. 1, with nearly square shapes of both
electron and hole bands responsible for the enhanced pairing.
These results seem to challenge an early view that $T_c$ is optimized
when a single band with $x^2-y^2$ symmetry crosses the Fermi surface
\cite{Oht91,Pav01,Sak10}, see also section 2.

Another way of increasing the hole density in Cu($3d$) orbitals was
realized for a CuO$_2$ monolayer grown on Bi2212 cuprate \cite{Zho16}.
Here the CuO$_2$ monolayer is
heavily overdoped by charge transfer at the interface and has a short
bond between Cu and the apical O in the substrate. A minimal
two-orbital model predicts indeed a nodeless high-$T_c$ SC state with
$s^{\pm}$ pairing \cite{Jia18}, arising from the spin-orbital exchange
a'la Kugel-Khomskii model \cite{Kug82}.

\begin{figure*}[t!]
\centering
\includegraphics[width=16.2cm]{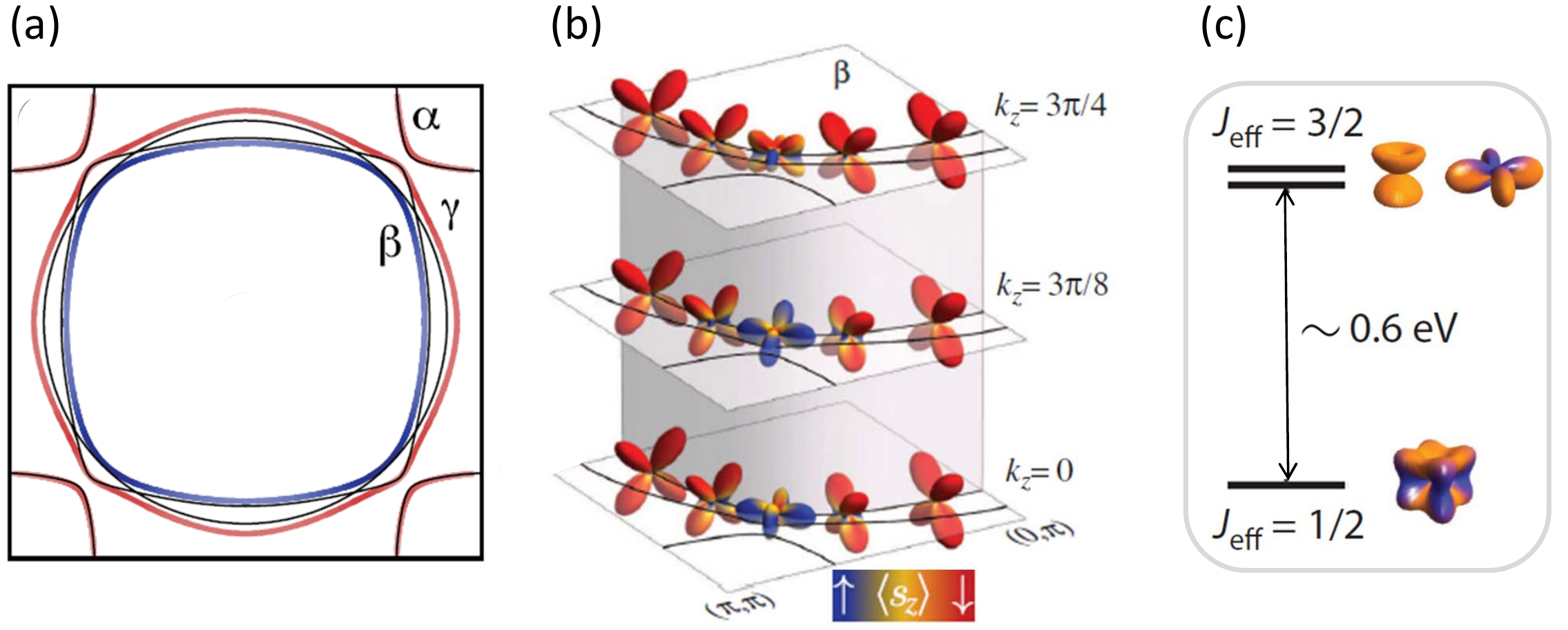}
\caption{(\textbf{a}) \& (\textbf{b})
Spin-orbital entanglement in Sr$_2$RuO$_4$ ruthenate:
(\textbf{a}) Fermi surface for bands $\{\alpha,\beta,\gamma\}$ at
$k_z=0$ calculated without (thin black) and with (thick, color-coded)
spin-orbit $\langle\vec{l}\cdot\vec{s}\rangle$ coupling, and
(\textbf{b})~momentum-dependent Ru($4d$) orbital projection of the wave
function for the $\beta$ band at selected momentum locations along the
3D Fermi surface. The orbital color represents the momentum-dependent
spin $\langle s^z\rangle$ expectation value; blue/red correspond to
spin ${\uparrow}/{\downarrow}$ for one state of the Kramers-degenerate
pair, see the color scale at bottom right. The strongly mixed colors
indicate momentum-dependent spin-orbital entanglement.
Images (\textbf{a}) and (\textbf{b}) are reproduced from~\cite{Vee14}.
(\textbf{c})~Orbital excitations in Sr$_2$IrO$_4$ between the entangled
spin-orbital ground state (with $J_{\rm eff}=1/2$) and the excited
doublet (with $J_{\rm eff}=3/2$), with the energy much lower
than in cuprates (cf. Table 1).}
\label{fig:Ru}
\end{figure*}

\subsection{Nickelates}
\label{sec:Ni}

One expects that another route to achieve higher hole doping in
Cu($3d$) orbitals is realized in Ni oxides, and indeed the Fermi
surface very similar to that for the overdoped cuprates was observed in
Eu$_{2-x}$Sr$_x$NiO$_4$ \cite{Uch11}. This Fermi surface is also
similar to the one found for LaNiO$_3$/LaAlO$_3$ heterostructure, where
both $e_g$ symmetries contribute to the band structure obtained in
local density approximation (LDA) \cite{Han09}, see Fig. 2. Although it
was argued that the $3z^2-r^2$ symmetry would be eliminated by electron
correlations, both bands were observed in the experiment \cite{Uch11}.

The theoretical study of the electronic structure for the
LaNiO$_3$/LaAlO$_3$ heterostructure suggests that \textit{orbital
engineering} using heterostructuring is in principle possible. The
electronic structure obtained with LDA for the heterostructure without
strain, see Fig. 2(a), has the Fermi surface at $k_z=0$ very similar
to that of the overdoped CuO$_2$ monolayer, cf. Fig. 2(b) and the right
inset of Fig. 1. Hence, the theory predicts a SC heterostructure, if
this system could be synthesized in the future. So far, widely tunable
orbital configurations were indeed realized by in strongly correlated
systems: LaTiO$_3$/LaNiO$_3$/LaAlO$_3$ heterostructure \cite{Dis15},
and nickelate superlattices \cite{Wu13,Hep14}.

Also in other square-planar nickelate, Pr$_4$Ni$_3$O$_8$, both
$e_g$ orbitals contribute to the Fermi surface \cite{Zha17}. X-ray
absorption shows that low-spin configuration with $x^2-y^2$ character
of the hole states is realized there \cite{Bot17}, making these
states remarkably similar to those in hole-doped cuprates.
Hence, also these compounds may be considered as promising candidates
for unconventional superconductivity. Recently, a family of Ni-based
compounds, which contain [Ni$_2$M$_2$O]$^{2-}$ (M-chalcogen) layers
with an antiperovskite structure constructed by mixed-anion Ni
complexes, has been suggested as possible high-$T_c$ superconductors
\cite{Le18}. Here again both $e_g$ symmetries should contribute, and
one expects strong competition between $s$-wave and $d$-wave pairing
symmetries.

More complicated situations are also possible, and we mention here
KNi$_2$S$_2$ as an example of a Ni-based superconductor with three
different orbital symmetries contributing at the Fermi surface and a
small value of $T_c=0.46$ K \cite{Lu13}. The electronic structure is
described by the multiorbital Hubbard model and this system has
certain similarity to iron-based superconductors discussed in
section 3.4.

\subsection{Superconducting ruthenate Sr$_2$RuO$_4$}
\label{sec:Ru}

The SC state of strontium ruthenate Sr$_2$RuO$_4$ was discovered in
1994 by Maeno and his collaborators after they had succeeded in
synthesizing high-quality samples of the material \cite{Mae94}. It was
soon realized that different orbital symmetries contribute to the SC
state in this unique ruthenium oxide \cite{Agt97}.
The value of $T_c=1.5$ K is not particularly high, but the ruthenate
structure suggests a possible relationship to the high-$T_c$ cuprate
superconductors. Yet, in spite of their structural similarity, doped
La$_2$CuO$_4$ and Sr$_2$RuO$_4$ are quite different \cite{Mae01}.

In conventional superconductors, in high-$T_c$ copper oxides, and in
Fe pnictides, the Cooper pairs have even parity. In contrast,
Sr$_2$RuO$_4$ is the candidate for odd parity superconductivity
\cite{Mac03}. However, the pairing symmetry in this material is not
yet fully established (see, e.g. \cite{Sud09,Pus19}). Ruthenates have
active $t_{2g}$ orbital degrees of freedom \cite{Wan13,Ima13}, with
degenerate $yz$ and $zx$ orbital states. An additional complication in
the theory is the presence of a sizeable spin-orbit coupling
($\zeta\sim 90$ meV at the $\Gamma$ point), which is much smaller than
full bandwidth of $\sim 3$ eV but nonetheless leads to important
consequences by mixing the low-energy $t_{2g}$ spin-orbital states. As
a result, the Fermi surface, consisting of three $\{\alpha,\beta,\gamma\}$
bands \cite{Ake19}, changes qualitatively, see Fig. 3(a).
In particular, the crossing points (accidental degeneracies) between
the electronic $\beta$ and $\gamma$ bands vanish and the Fermi surface
for both $\{\beta,\gamma\}$ bands becomes more circular \cite{Vee14},
while there is no phase space for mixing of hole-like $\alpha$ band.

For further discussion we focus on the representative $\beta$ band; the
mixing of $t_{2g}$ orbital states remains similar for the other bands.
The most important consequence of finite spin-orbit coupling $\zeta$ is
momentum-dependent spin-orbital entanglement \cite{Vee14} of the
eigenstates near the Fermi surface. It is illustrated in Fig. 3(b) for
the $\beta$ band and three representative values of momentum $k_z$.
Going along the
Fermi surface for a fixed value of $k_z$, one observes that not only
the orbital character changes, but also both components of $s=1/2$
spin mix. The latter mixing of spin components is somewhat weaker for
large value of $k_z=3\pi/4$. This mixing plays an important role and
the challenging theoretical problem is to include orbital fluctuations
in the theory of superconductivity in Sr$_2$RuO$_4$.

When spin-orbit coupling is large enough, it can unify the spin and
orbital subspaces locally, forming the effective pseudospin
$J_{\rm eff}=1/2$ states that dominate low energy physics
\cite{Kha04,Kha05,Jac09}. This case is particulary relevant for
iridium oxides, and indeed, a perovskite compound Sr$_2$IrO$_4$ was
found to host pseudospin $J_{\rm eff}=1/2$ antiferromagnetism
\cite{Kim08,Kim09,Kim12}, with quasi-2D magnon excitations similar to
those of spin $S=1/2$ in cuprates.
It was recently found that spin-orbit entangled magnetism in iridates
and ruthenates is strongly influenced by
electron-lattice coupling, via pseudo-Jahn-Teller effect \cite{Liu19}.
The remarkable analogy between layered iridates and cuprates holds
also upon doping --- single band Fermi surface, (pseudo)gap,
and Fermi arcs have been detected \cite{Kim14}. However, a long-range
coherent superconductivity has not been yet found in iridates. Whether
this is related to the fact that spin-orbital excitation energies 0.6
eV \cite{Kim12} in iridates, see Fig. 3(c), are much lower than in
cuprates, or to the Mott insulating nature of iridates, in contrast
to charge-transfer insulating cuprates, remains an open question. For
detailed discussion the similarities and differences between iridates
and cuprates, we refer to the recent review article \cite{Ber19}.

\subsection{Iron-based superconductors}
\label{sec:Fe}

Similar to cuprates, the Fe-based superconductors have 2D lattices of
$3d$ transition metal ions as building blocks. However, while oxygen
ions lie in the same planes as Cu ions in La$_2$CuO$_4$, ions of As, P,
or Se lie above or below the Fe plane, in positions close to
tetrahedral. For this reason, the out-of-plane As
orbitals hybridize well with $t_{2g}$ orbitals of Fe ions. In addition,
there is a substantial overlap between the $3d$ orbitals
\cite{Pag10,Hir11,Sca12,Si16,Fer17,Ima05,Boe11,Roser}. Under these
circumstances, the minimal models for pnictide superconductors contain
at least two \cite{Raghu} or three \cite{Dag10} $t_{2g}$ orbitals per
Fe atom.

In contrast to undoped cuprates, parent compounds of Fe-based
superconductors are metallic \cite{Pag10}, and the pairings of
different symmetry compete in a two-band model
\cite{Fer17,Fer10,Nic11,Nic12}. The coexistence of hole-like and
electron-like bands at the Fermi surface is quite generic
\cite{Pag10,Hir11,Boe11}. Then Lifshitz transitions can develop for
increasing external magnetic field. Such transitions were indeed found
in a two-band model with intra-band pairing \cite{Pto17} and could
explain the experimental observations in FeSe and Co-doped
BaFe$_2$As$_2$ compounds.
The competition between different pairing symmetries in KFe$_2$As$_2$
causes a change of symmetry at the critical pressure, from $d$-wave to
$s_{++}$-wave symmetry \cite{Taf13}.

Before discussing the nature of pairing, it is interesting to look at
the strong coupling model for pnictides, including magnetic
frustration \cite{Qi08}. A spin-orbital model for interacting Fe ions in
intermediate $S=1$ spin states was derived \cite{Zaa09} in the regime
of strong electron correlations.
It highlights Hund's exchange for electron correlations, recently
observed experimentally \cite{Fink}. Magnetic and orbital instabilities
are here far richer than in cuprates, and one finds the experimentally
observed spin-stripe state which could be accompanied by three
different types of orbital order. This is another manifestation of
substantial magnetic frustration of the superexchange \cite{Qi08} which
may partly explain why magnetic instabilities are sometimes absent,
e.g. in FeSe and LiFeAs.

The spin-orbital model \cite{Zaa09}, however, does not contain
biquadratic exchange which was found to be crucial in magnetism of
Fe-based superconductors \cite{Yar09,Wys11,Yu12,Val15}. This coupling
may originate from spin-state fluctuations \cite{Cha13} typical
for Fe-ions in systems with strong $p-d$ covalency. While large
biquadratic exchange is quite unique and essential for the description
of magnetic and nematic instabilities, its
implications for SC instabilities belong to open problems in the field.
Indeed, the SC phase in iron-based materials occurs in the vicinity of
the two above instabilities: Not only one has the usual magnetic phases
\cite{Dai12}, but also the nematic order may occur
\cite{Sch12,Bea14,Wat15}. A detailed study shows that the Fermi surface
of FeSe undergoes a spontaneous distortion from fourfold-symmetric to
twofold-symmetric elliptical pockets, and next SC phase emerges from
the nematic electronic phase.

The theory of Cooper pairing in Fe-based superconductors is rather
involved and still far from complete \cite{Hir11,Sca12,Fer12}.
The role played by orbital and spin fluctuations
belongs to challenging open problems in the theory. It was
found that the orbital fluctuations may give rise to the strong
pairing interaction due to the cooperation of Coulomb and
electron-phonon interactions \cite{Kon10,Ona12}. The theory explains
also the famous empirical relation between $T_c$ and the As-Fe-As bond
angle \cite{Sai10}.

Altogether, the pairing in iron-based superconductors involves all the
five Fe($3d$) orbitals, and multiple orbital physics gives rise to
various novel phenomena like orbital-selective Mott transition,
nematicity, and orbital fluctuations that may support the SC phase.
Recent theory treating spin and orbital fluctuations on equal footing
predicts that, at certain conditions, a spontaneous orbital order sets
in first, and then superconductivity follows \cite{Chu16}. The SC gap
and low energy excitations in FeSe are dominated by a single $xz$
orbital \cite{Liu18} which uncovers the orbital origin of the strongly
anisotropic gap in the FeSe superconductor
\cite{Has18,Spr17,Nic17,Ptok}.
In LiFe$_{1-x}$Co$_x$As spin excitations are orbital selective:
low-energy spin excitations are mostly from $xy$ orbitals, while
high-energy spin excitations arise from the $yz$ and $zx$ orbitals
\cite{Li16}. Such strongly orbital selective spin excitations in LiFeAs
family might play a role in the mechanism of orbital selective Cooper
pairing as well.

\section{Summary}
\label{sec:summa}

To conclude, we have presented the current status of the high-$T_c$
superconductivity in the presence of orbital degrees of freedom.
As this subject is very broad, in this review we limit the
presentation solely to the electronic degrees of freedom, leaving
aside a detailed discussion of the role played by their coupling to
the lattice.
In case of cuprates large hole doping and removing octahedral
distortions is necessary to activate the $3z^2-r^2$ orbitals, and we
gave examples of cuprates with two $e_g$ orbitals and higher values
of $T_c$ than in La$_{2-x}$(Sr,Ba)$_x$CuO$_4$. These orbitals
contribute to almost the same Fermi surface, consisting of hole and
electron parts, also in doped nickelates. But for nickelates we cannot
present anything more than a theoretical suggestion that the
superconducting instabilities could occur as well.
Finally, very interesting superconducting states emerge in
Sr$_2$RuO$_4$ and in iron pnictides, where several $t_{2g}$ symmetries
meet at the Fermi surface and participate in the Cooper pairing.
Search for other transition metal compounds with SC instabilities
continues --- for instance, recently AgF$_2$ was suggested as an
excellent analogue to La$_2$CuO$_4$
\cite{Gaw19}, but no superconductivity was observed so far.

\begin{quote}
Summarizing, the presence of orbital degrees of freedom makes
high-$T_c$ superconductors an even more exciting class of quantum
materials where the competing quantum phases are of particular
importance for superconductivity in layered compounds \cite{Jar19}.
It seems that orbital fluctuations could enhance the superconducting
transition temperature $T_c$, but we emphasize that the role of orbital
degrees of freedom in the phenomenon of pairing belongs to open
problems in the theory; in particular the interplay between orbital
degeneracy and the Jahn-Teller coupling to lattice --- the idea that
guided Bednorz and M\"uller in their discovery of high-$T_c$
superconductivity --- has to be worked out in a greater detail.
\end{quote}

\textbf{Authors' Contributions:} All authors selected the relevant
information, participated in discussions, wrote the manuscript and
contributed to the interpretation of the results.

\textbf{Funding:} This research was funded by Narodowe Centrum Nauki
(NCN, National Science Centre, Poland) under Projects
Nos. 2016/22/E/ST3/00560 and 2016/23/B/ST3/00839.

\textbf{Acknowledgments:}
We would like to thank Mona Berciu, Antonio Bianconi, Lucio Braicovich,
Jeroen van den Brink, Mario Cuoco, Maria Daghofer, Thomas P. Devereaux,
Louis Felix Feiner, J\"org Fink, Andres Greco, Maurits Haverkort, Peter
J. Hirschfeld, Peter Horsch, Liviu Hozoi, Huimei Liu, Andrzej Ptok,
Roman Pu\'zniak, George A. Sawatzky, J\'ozef Spa\l{}ek, Hiroyuki Yamase,
Alexander N. Yaresko, Jan Zaanen, and Roland Zeyher for many insightful
discussions.
A.~M.~Ole\'s is grateful for the Alexander von Humboldt
Foundation Fellowship (Humboldt-Forschungspreis).

\textbf{Conflicts of Interest:} The authors declare no conflicts of interest.

\end{document}